# Hierarchical hydropathic evolution of influenza glycoproteins (N2, H3, A/H3N2) under relentless vaccination pressure

J. C. Phillips

Dept. of Physics and Astronomy, Rutgers University, Piscataway, N. J., 08854

1-908-273-8218 jcphillips8@comcast.net

## Abstract

Hemagglutinin (HA) and neuraminidase (NA) are highly variable envelope glycoproteins. Here hydropathic analysis, previously applied to quantify common flu (H1N1) evolution (1934-), is applied to the evolution of less common but more virulent (avian derived) H3N2 (1968-), beginning with N2. Whereas N1 exhibited opposing migration and vaccination pressures, the dominant N2 trend is due to vaccination, with only secondary migration interactions. Separation and evaluation of these effects is made possible by the use of two distinct hydropathic scales representing first-order and second-order thermodynamic interactions. The evolutions of H1 and H3 are more complex, with larger competing migration and vaccination effects. The linkages of H3 and N2 evolutionary trends are examined on two modular length scales, medium (glycosidic) and large (corresponding to sialic acid interactions). The hierarchical hydropathic results complement and greatly extend advanced phylogenetic results obtained from similarity studies. They exhibit simple quantitative trends that can be transferred to engineer oncolytic properties of other viral proteins to treat recalcitrant cancers.

## Introduction

Earlier papers on the panoramic smoothing evolution of H1N1 glycoproteins showed that several dominant factors could be quantified by hydropathic analysis of the interactions between amino acid (aa) sequences and the water film that envelopes or packages them [1-3]. Prior to 1950, this evolution exhibited only small fluctuations due to direct antibody evasion. However, after the introduction of widespread vaccinations by the US Army in 1944, evolution after 1950 was dominated by vaccination evasion, which progressively smoothes the water-protein interface, and opposing migration pressures, which compact the protein and regressively increase the roughness of the water-protein interface. These opposing interactions accumulate and lead to Darwinian punctuations. Here we extend the analysis to H3N2, in order to develop further insights into smoothing mechanisms, with a view to engineering viral oncolytics.

There are several extensive studies of the genetic evolution of H3N2 based on counting individual aa mutations, separated according to whether they are synonomous or not (described as positive in BLAST) [4]. Similarities correlate geographically and temporally with local clustering effects measured indirectly (antigenically with antibody titres) [5]. On a panoramic scale these clustering effects are small compared to the fundamental genetic hydropathic trends of H1N1 [1-3] and H3N2 (discussed here)

There are at least two much larger fundamental hydropathic effects. The first depends on the length scales of native glycosidic or sialic acid interactions, while the second depends on whether the mutations are thermodynamically nearly first order (treated by the KD scale) or second order (MZ scale). Here we analyze these two effects quantitatively, and show that they display a number of mechanisms that have previously been described for H3N2 by schematic models [6,7]. These models exhibit canalization, convergence, modular and punctuation effects that have larger implications for RNA transcriptive mechanisms [8,9]. In contrast to these general qualitative models, the present results include all these effects quantitatively, and do not rely on adjustable parameters. Hydropathic analysis can be used to engineer similar viruses with enhanced oncolytic properties.

## Methods

The methods used here are almost the same as were used in [1] for NA, but with a few important changes. The evolution of NA was monitored by calculating the roughnesses of $R_{KD}(W_{max})$ and $R_{MZ}(W_{max})$ with $W_{max}$ = 17 and 47 of the water packaging film with two scales, KD and MZ. The two length scales $W_{max}$ = 17 and 47 correspond respectively to membrane thickness and glycosidic spacing. These N1 roughnesses (especially with the MZ scale) showed well-defined plateaus connected by punctuated decreases (vaccination program successes) or increases (migration), culminating in vaccination-driven N1 convergence to a nearly common strain after the 2007 swine flu program. Here punctuation of N2 is not visible on our larger scale (although it might be visible in an elaborate statistical analysis that combines BLAST similarities [4,5] with our hierarchies). Instead we see canalization with the first-order KD scale, and migration fluctuations with the second-order MZ scale. These differences between H1N1 and H3N2 can be explained by the different origins of H1N1 (primarily human) and H3N2 (frequent intermixing with avian) strains [6].

## Results (N2)

To obtain an unbiased survey with fewer strains, data were obtained from three sites, Memphis, Hong Kong, and the Netherlands, with significantly different climates. These have led to large geographically dependent and sparsely measured antigenic differences, as measured on ferret post-infection inhibition titres [6]. These antigenic differences [5] may be reflected in the large scatters seen in the $R_{MZ}(W_{max})$ with $W_{max}$ = 17 and 47 data (MZ scale) shown in Figs. 1 and 2. Unexpectedly this scatter disappears in $R_{KD}(W_{max})$ with $W_{max}$ = 17 and 47, as shown in Figs. 3 and 4 (KD scale).

One of the largest scatters occurs between $R_{MZ}(17 \text{ or } 47)$ for Mem and HK 1972. Inspection of the corresponding hydroprofiles shows that this difference is caused by a single philic - phobic mutation, N336Y. Another large $R_{MZ}(17 \text{ or } 47)$ difference occurs between Mem and Neth 1999, which is traced to two mutations, L370S (phobic-philic) and S414G (philic-neutral). The collapse of $R_{MZ}(17,47)$ onto $R_{KD}(17,47)$ occurs because of differences between the MZ and KD values [10] for $\psi(Y)$ in the 1972 example, and $\psi(S)$ in the 1999 example. For the reader's convenience, the MZ and KD scales are shown in Table I, normalized to their average values. The largest effects on roughness occur for a large change in $\psi(i)$ that occurs near a peak or valley in $\psi(i,W)$. An example is shown in Fig. 5 for N336Y of Mem and HK 1972.

At first one could suppose that this MZ-KD collapse is accidental, but as it occurs for the entire panorama 1968- and for three different sites with different climates, this is most unlikely. A more plausible explanation is that the KD and MZ scales do in fact describe first-order and second-order thermodynamic differences, as implied by their origins [1-3]. Common flu H1N1 has evolved over perhaps 500 or more years, and has approached a critical point, first through antibody evasion, and more recently through vaccination pressures (since 1945). The N1 panorama exhibits large (but still second-order) migration perturbations in its decreasing roughness. By contrast, here N2 roughness decreases steadily (1968-) using the KD scale. The N2 subtype has emerged by reassorting human and avian strains [6], and an initially rapid roughness descent would be better described by the first-order KD scale than by the second-order MZ scale that worked well for N1 and H1N1. The internal consistency of this picture provides support for distinguishing hydropathicity scales in thermodynamic terms.

The collapse of broadened MZ panoramic evolution curves onto channeled KD curves enables us to examine the latter in Fig. 4 for distinctive features, which are readily obvious. Widespread vaccination for H3N2 began in 1968 with a Hong Kong strain [6], which was replaced by an English strain in 1972. Substantial decreases in KD roughness appear for all three sites between 1972 and 1975, with the effects being much larger for W = 47 (Fig. 4) than for W = 17 (Fig. 3).

From 1975 to present, $R_{KD}(17)$ decreases nearly linearly by more than 20%, echoing a similar decrease for N1 of H1N1. Similarly $R_{KD}(47)$ decreases, but much more rapidly and steeply by more than 35% up to 1995, when it abruptly levels off. This larger decrease and leveling after 1995 suggests that vaccination pressure on H3N2 affects the glycosidic length scale W = 47 well before it affects the membrane length scale W = 17. This larger effect on W = 47 is not a surprising result, as we already saw [2] that H1 and N1 are strongly coupled, and that the dominant interactions in H1 occur on the even larger length scale of W =111, appropriate to sialic acid interactions. These longest range interactions are optimized first, while the weaker medium-range glycosidic interactions are apparently still being optimized at the W = 17 level in 2009, but could saturate soon (5-10 years).

The overall picture for N1 in H1N1 and N2 in H3N2 is that both glycoproteins have evolved under vaccination pressure to nearly minimal roughness. As discussed before, the panoramic evolution of N1 roughness has occurred in tandem with infection severity, as reflected most recently in the numbers of strains sequenced during the successive swine flu outbreaks (New York 2003, Berlin 2005, Houston 2007). Current strains are more convergent, as expected from SOC, and are historically the mildest. So long as widespread vaccination programs are in place, this stabile mildness should persist in relatively uncrowded areas like America and Europe.

**Results (H3)**

The evolution of H1 roughness is qualitatively different from N1 evolution, but the two are linked [2,4]. The simple reductions in roughness of N1 and N2 are replaced by large block shifts (called proteinquakes) in the hydropathic profiles of H1 at the times when N1 is punctuated. H1 roughness decreases with time. With large fluctuations for H3N2, we find that for the HA1 chain H3 R(W) simply oscillates, with two large peaks in 1976 and around 2000 for W = 111 (Fig. 6(a)). We can resolve the broad 2000 peak by reducing W to 47, as shown in Fig. 6(b), which shows that the peak occurs in 1999, and is sharper with the MZ scale, with an abrupt increase between 1995 and 1999.

We can identify the cause of these oscillations by studying the MZ 47 hydroprofiles of H3. For H1 we already saw in Fig. 5 of [2] evidence for a proteinquake in 1976 associated with the sialic acid binding site. Here H3 has been channeled, which leads us to suspect that H3 could be hydrorigid, and exhibit much larger proteinquakes. Fig. 7 shows the HA1 chain profiles (W = 47) between the 1976 and 1999 peaks: there are very large H3 proteinquakes. The most striking shift occurs between 1985 and 1999, as a deep hydrophilic valley or hydrofault appears between 140 and 160 near the center of the HA1 chain. The 1985-1999, 140-160 hydrofault occurs because of multiple softening mutations [I137N, G140S, Y153S, and VN(160,161)NK]). The strengthening of the largest hydrophobic peak near 260 is caused by a single mutation, S263C.

How did vaccination pressure return Memphis 1999 R(47and 111) to the background level by 2008 (Fig. 6)? The 2008 high-resolution W = 47 hydroprofile is shown in Fig. 8. There is a large hydrophobic contraction between 130 and 230, which corresponds very well to the sialic acid binding region of H1, previously identified in Fig. 2 of [2], and similarly labeled here. Note that this region was previously identified in H1 by connecting it to structural data, but here the H3 identification is based solely on similarity of H3 proteinquakes to H1 proteinquakes.

There is a striking difference between the Memphis 1999 R(47and 111) profiles (Figs. 7,8 and 9), corresponding to glycosidic and sialic acid interactions, respectively. The deep double R(47) hydrofault from 140 to 170 in 1999 narrows and shifts to 170 -190 in 1999. It would be worthwhile to study linear epitopes on H3 by peptide scanning using libraries of overlapping peptides against convalescent sera from H3N2 patients similar to those done for H1N1 [12] and H5N1 [13].

What are the individual amino acid mutations responsible for this hydrophobic 1999-2008 shift? There are four hydrophilic to hydroneutral mutations, K166I, K171N, K189Q, S202G, and one hydrophilic to hydrophobic, S209F. The shift is cut off at by W237R, hydrophobic to hydrophilic, which reduces the largest hydrophobic peak near 260.

Clearly ionic interactions involving Lysine (K) in 1999 have been suppressed in 2008 in favor of hydrophobic compaction. Most of the sequences very similar to Memphis 1999 are earlier, and the nearest later ones are 2002 New York. Apart from hydrophilic to hydroneutral mutations, another way of suppressing charge interactions is to replace charged NK(160,161) in 1999 Memphis by neutral DK(160,161) in 2002 New York, where the two charges neutralize each other. If one restricts one's attention to epitopic subsets defined by Euclidean contacts at active sites and counts only total numbers of charged amino acids, including those in neutralized pairs, replacing NK by DK increases the number of charged mutations [14]. Given the overall consistency of the hydropathic approach, it appears that short-range ionic effects may be only a weaker secondary reaction to dominant long-range hydropathic elastic interactions, especially in determining sialic acid interactions.

## Discussion

One could ask which glycoprotein, H3 or N2, plays a larger role in evolution of virulence of H3N2. At present the answer would appear to be N2. The roughening of H3 in 1999 Memphis because of the appearance of the 140-170 R(47) hydrofault does not seem to have alarmed virologists, as there were few H3 sequences in Memphis in the period 1995-2004, and most of these appeared in a single year (2003), probably incidentally. By 1995 N2 had reached its low level of roughness, and this may explain H3N2 stable and low virulence after 1995. By 1999 the rotating WHO recommended H3N2 vaccine composition was Moscow/10 (AFM72208), very similar to Hong Kong 1999 and similar to New York 1999-1997. The hydroprofile of this strain also resembles that of Memphis 2003.

There is an interesting glass network analogy for the hydrophobic proteinquake of H3 between 1999 and 2008. Hydrophobic interfaces often dimerize [15], and one can suppose that a necessary plastic condition for such dimerization is that the molecular network interfaces involved are both nearly rigid and yet stress-free. Such networks have been widely observed in glass alloys composed of nearly isovolume chalcogenides, which are good inorganic analogues of protein amino acids (also nearly isovolume). A striking feature of these elastically optimized inorganic alloys is that their glass transitions are nearly reversible [16], a necessary condition for protein functionality, here involving dimerization.

Emergence of new strains can be detected by similarity tools using dynamical clustering [17]. If such a new strain should show a large uptick in N2 R(47), then its hydropathic clustering properties should be studied with care, as it might become dangerous.

**Conclusions**

Comparison of the present H3N2 results with those for H1N1 shows simpler smoothing for N2 here than for N1 in H1N1, confirming the widely held view of close H-N linking [4]. More interesting is the connection between canalization [7] and larger proteinquakes in H3. These are again dominated by sialic acid interactions, which lead to larger and more dramatic H3 proteinquakes than those previously identified for H1. The combined analysis of roughening evolution and hydroprofiles is most informative. It does not require new structural data, and it is extremely accurate in portraying the evolution of competing elastic and charge interactions.

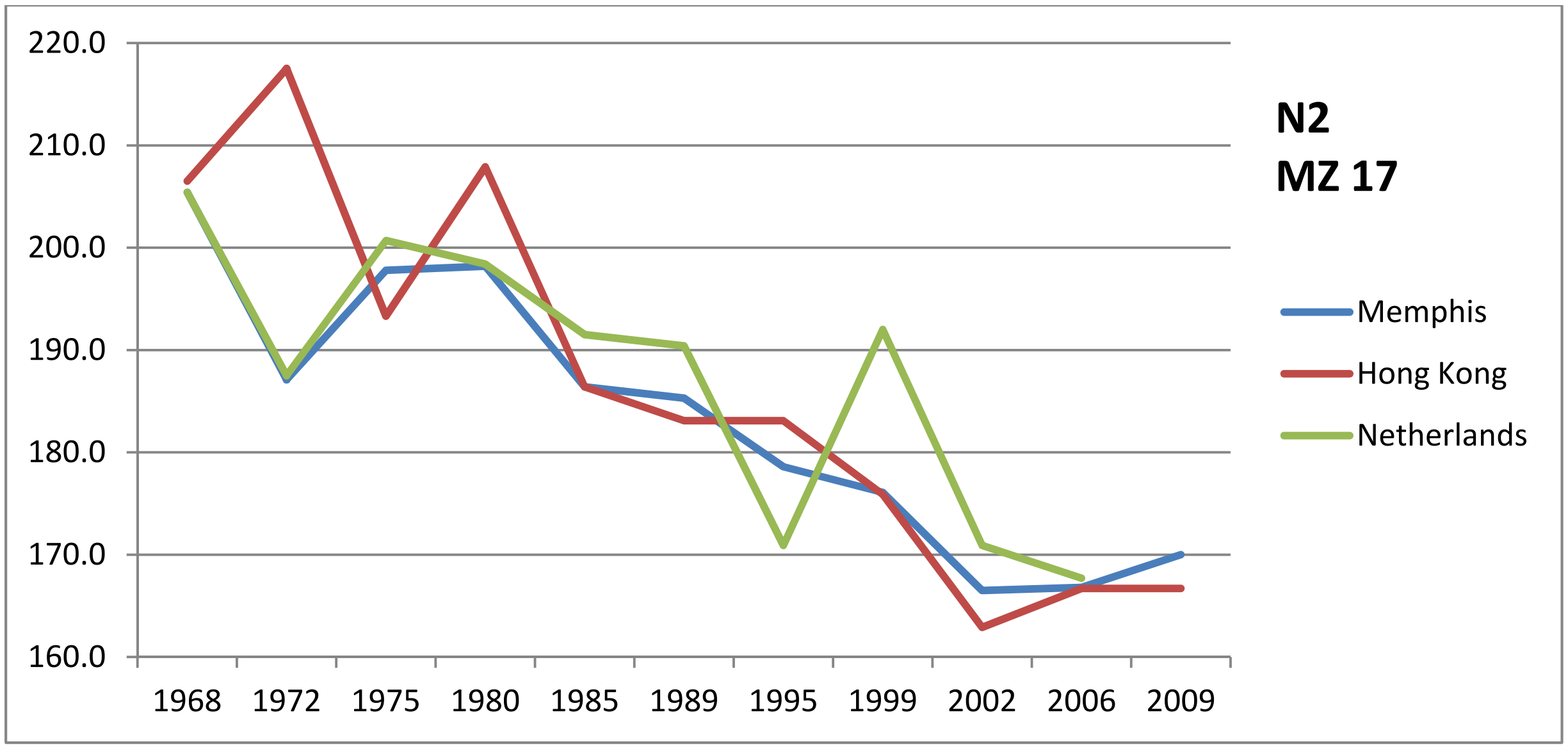

Fig. 1. Evolution of $R_{MZ}(17)$ at three sites with different climates and crowding conditions.

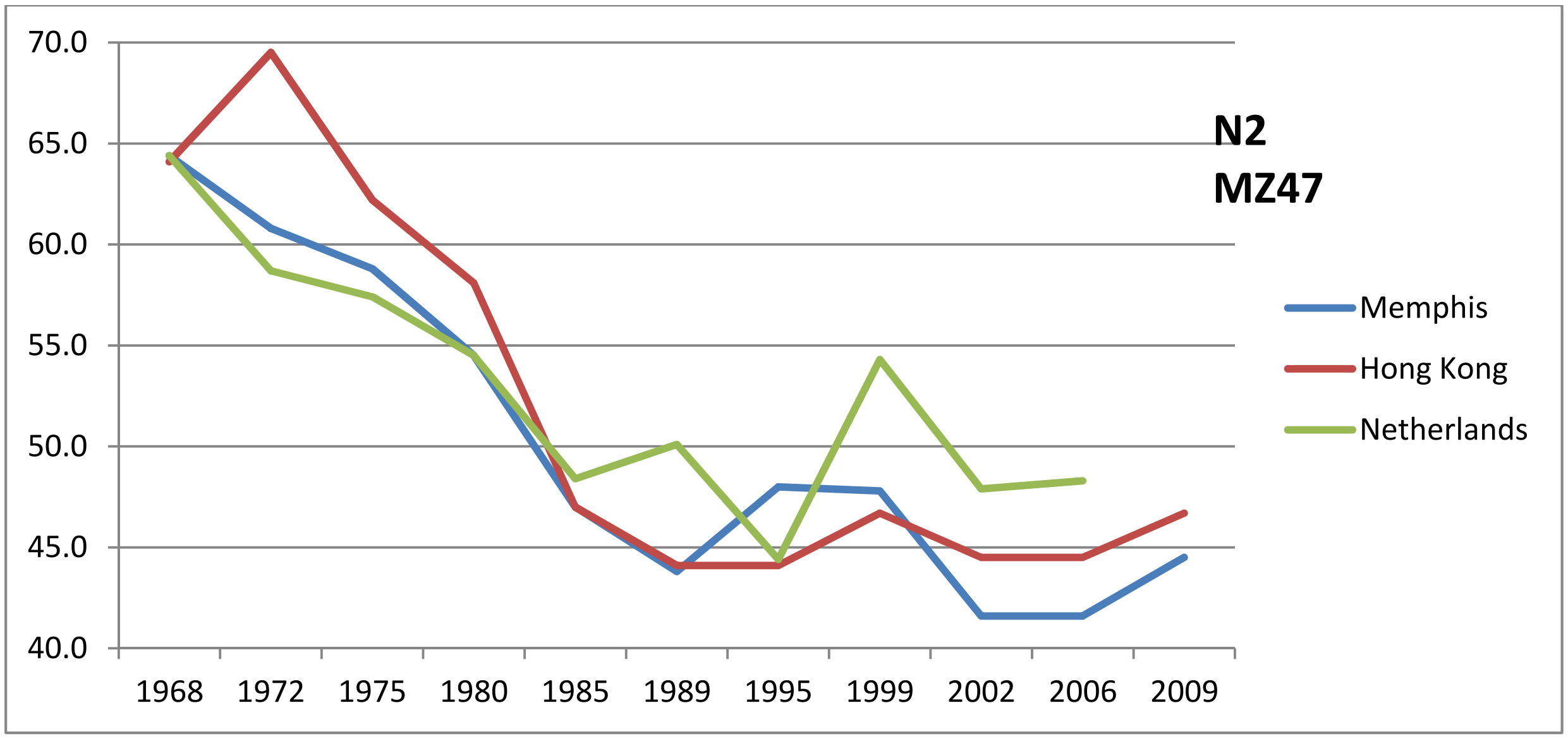

Fig. 2. Evolution of $R_{MZ}(47)$ at three sites with different climates and crowding conditions. The agreement is slightly improved.

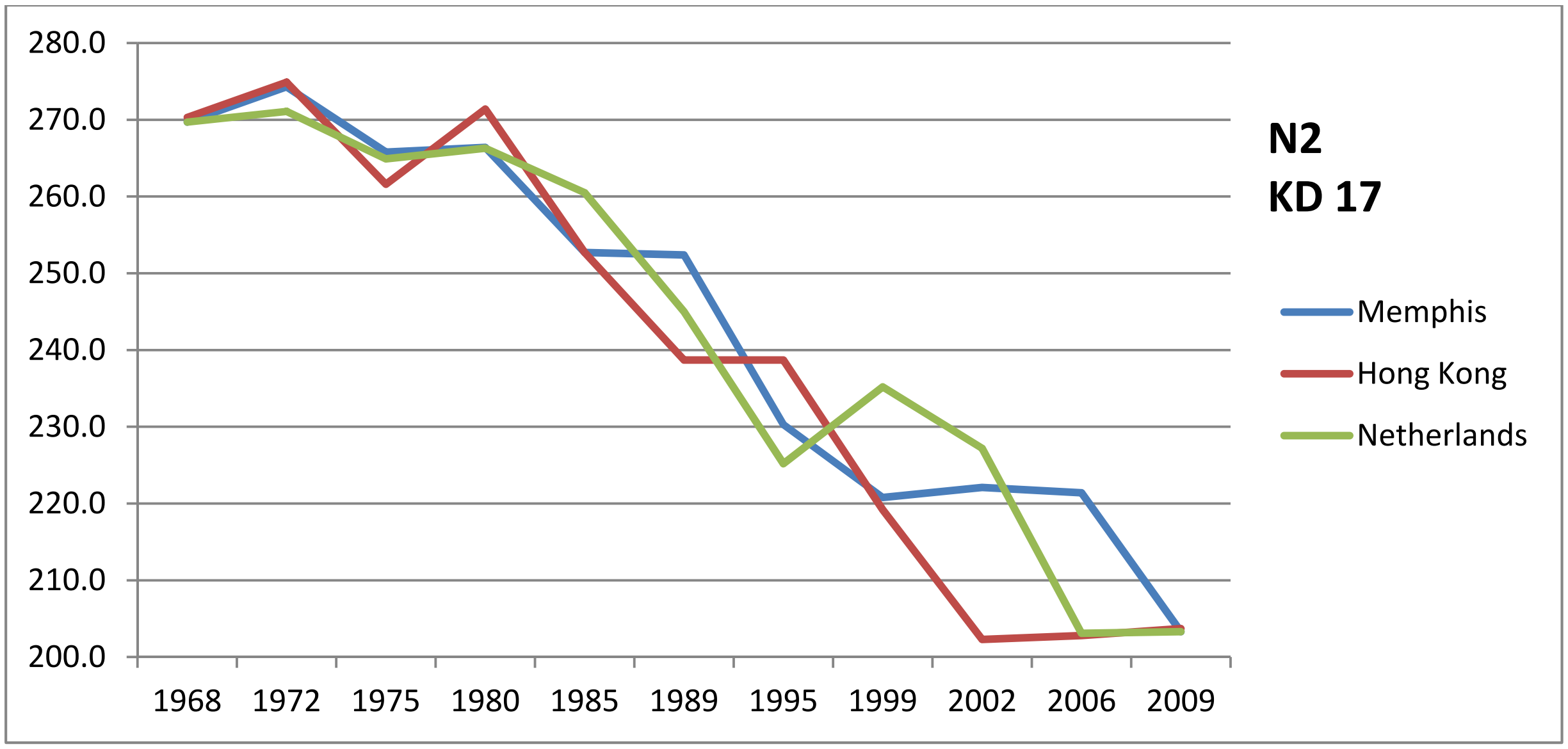

Fig. 3. Evolution of $R_{KD}(17)$ at three sites with different climates and crowding conditions. The use of the first-order KD scale, instead of the second-order MZ scale, improves concordance.

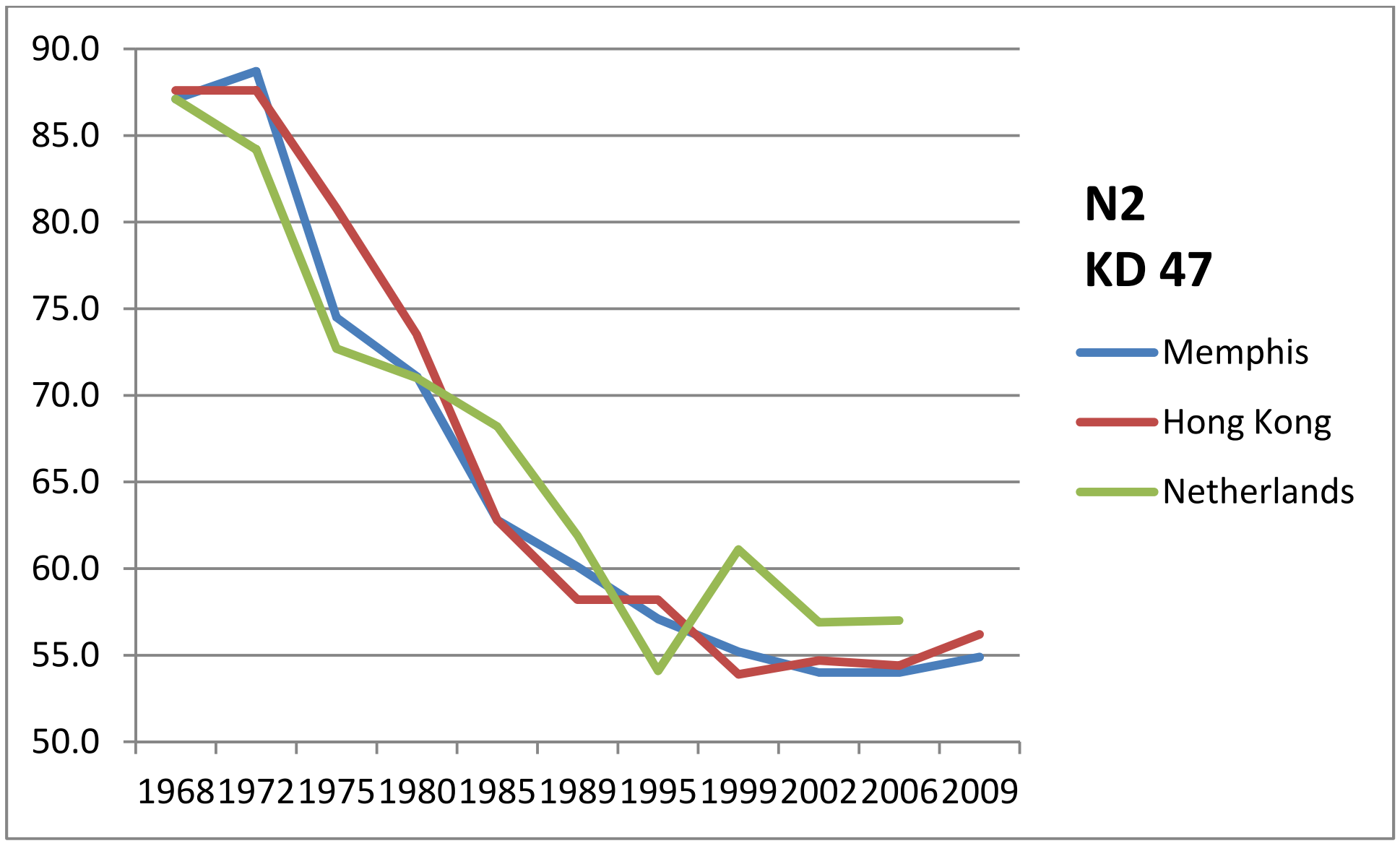

Fig.4. New evolutionary features (see text) are brought out here.

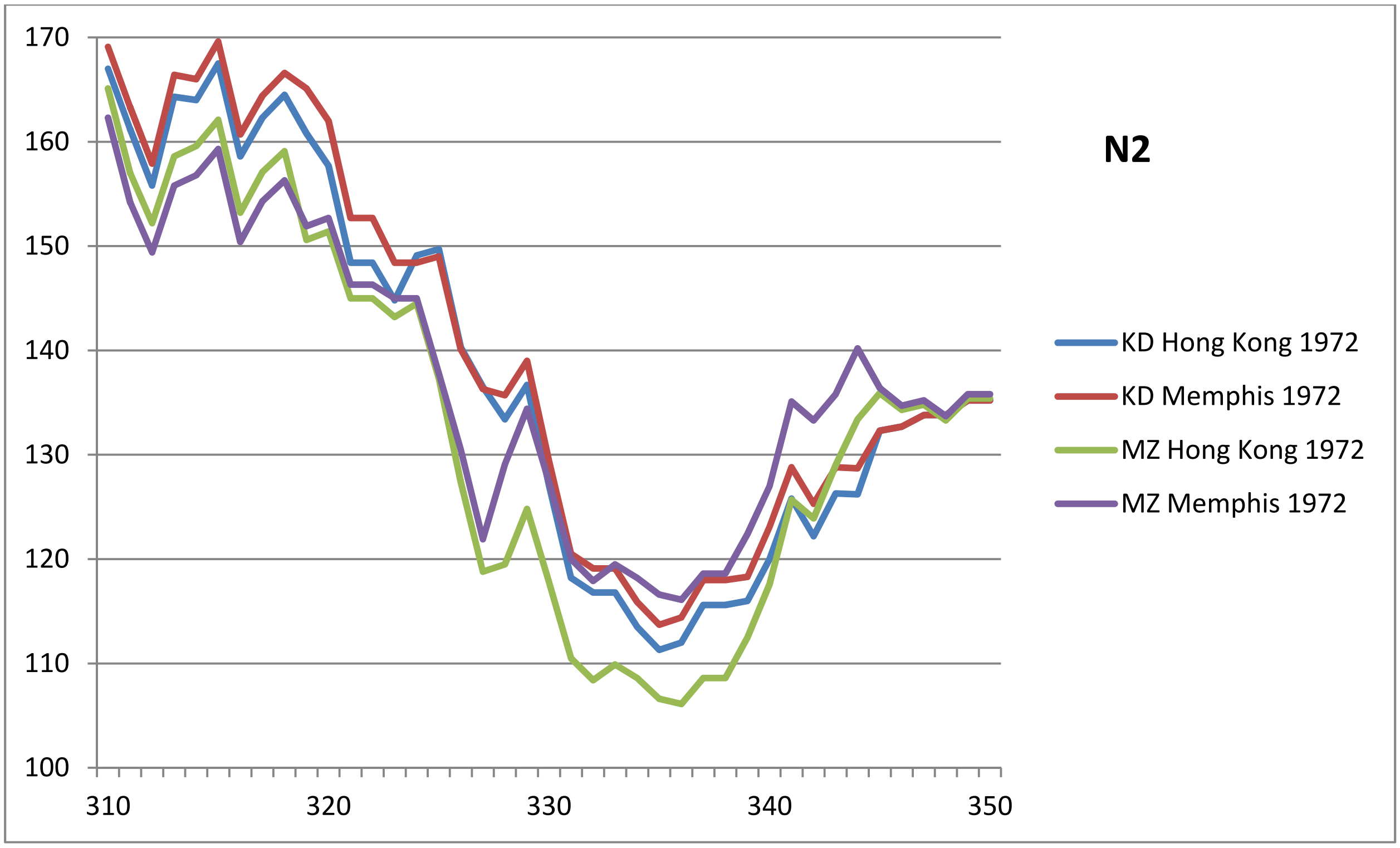

Fig. 5. The largest effects on roughness occur for a large change in ψ(i) that occurs near a peak or valley in ψ(I,W). An example is N336Y of Mem and HK 1972.

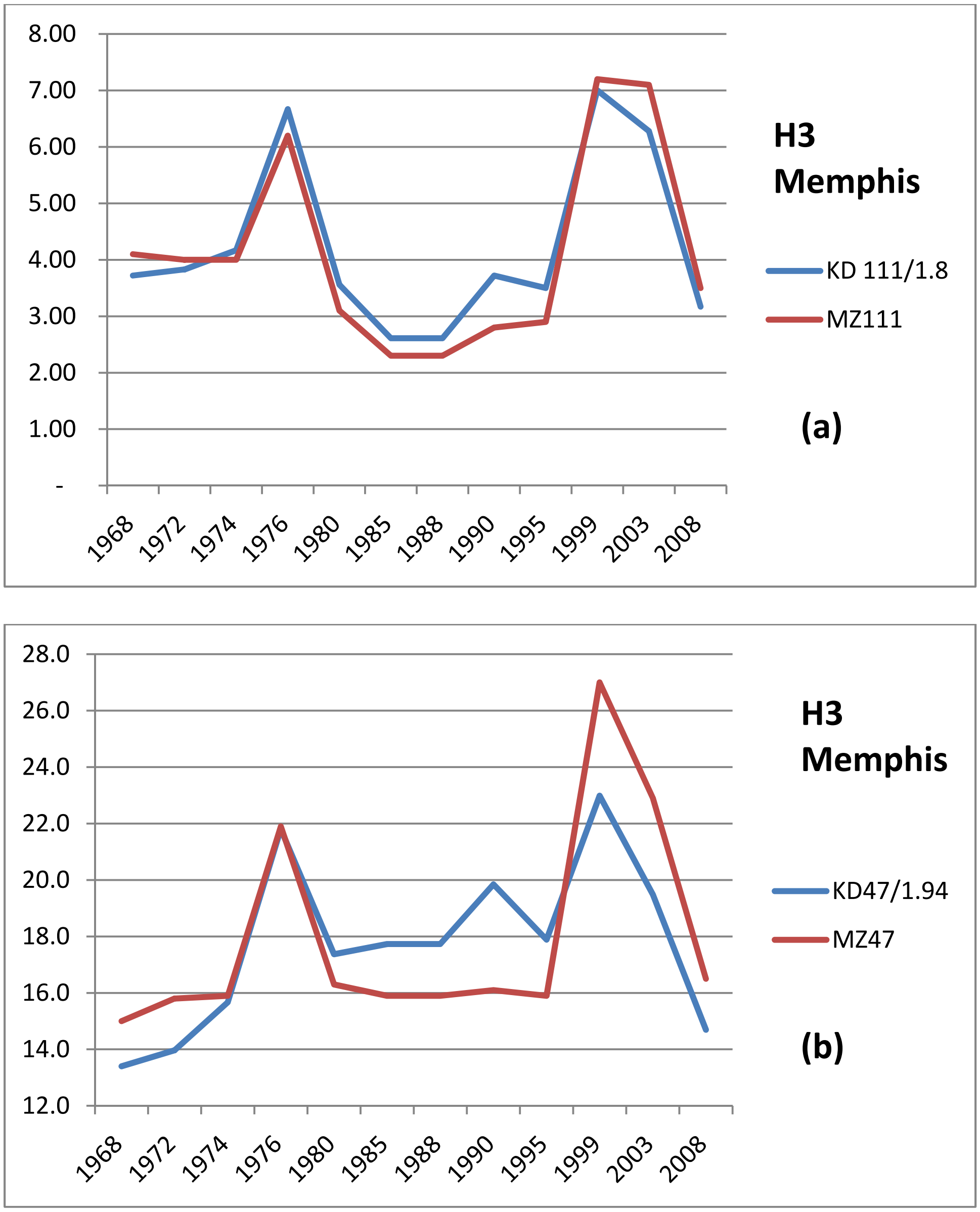

Fig. 6. (a) Evolution of $R_{KD}(111)$ and $R_{MZ}(111)$ for HA1 chain of H3 with W = 111, (b) Same at higher resolution with W = 47. Note how the 2001 doublet of (a) has been resolved.

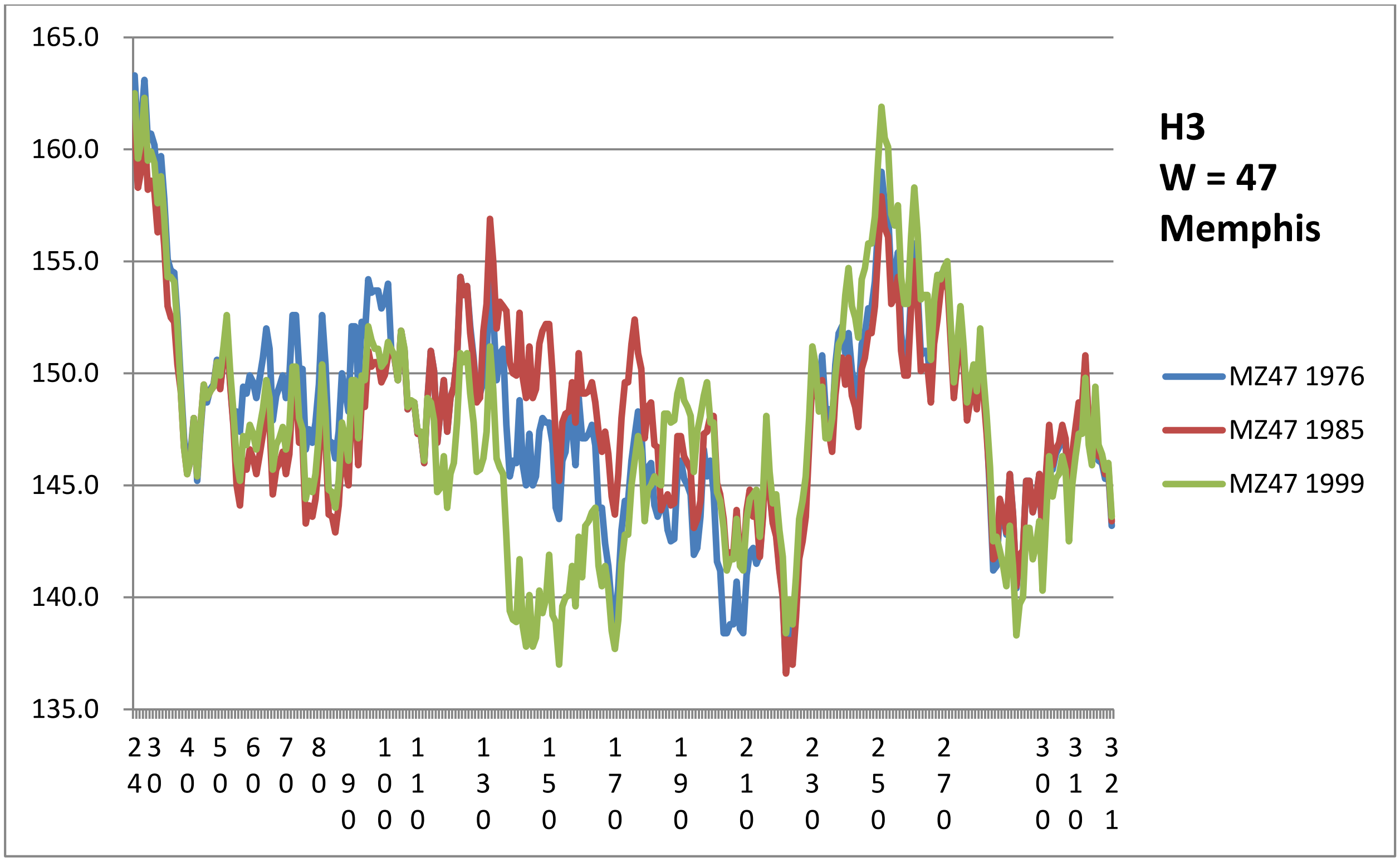

Fig. 7. Evolution of chain HA1 profiles of H3 from 1976 (ABD16740) to 1985 (ABD61777) to 1999 (AAZ43405) exhibits a deep double hydrofault from 140 to 170 in 1999.

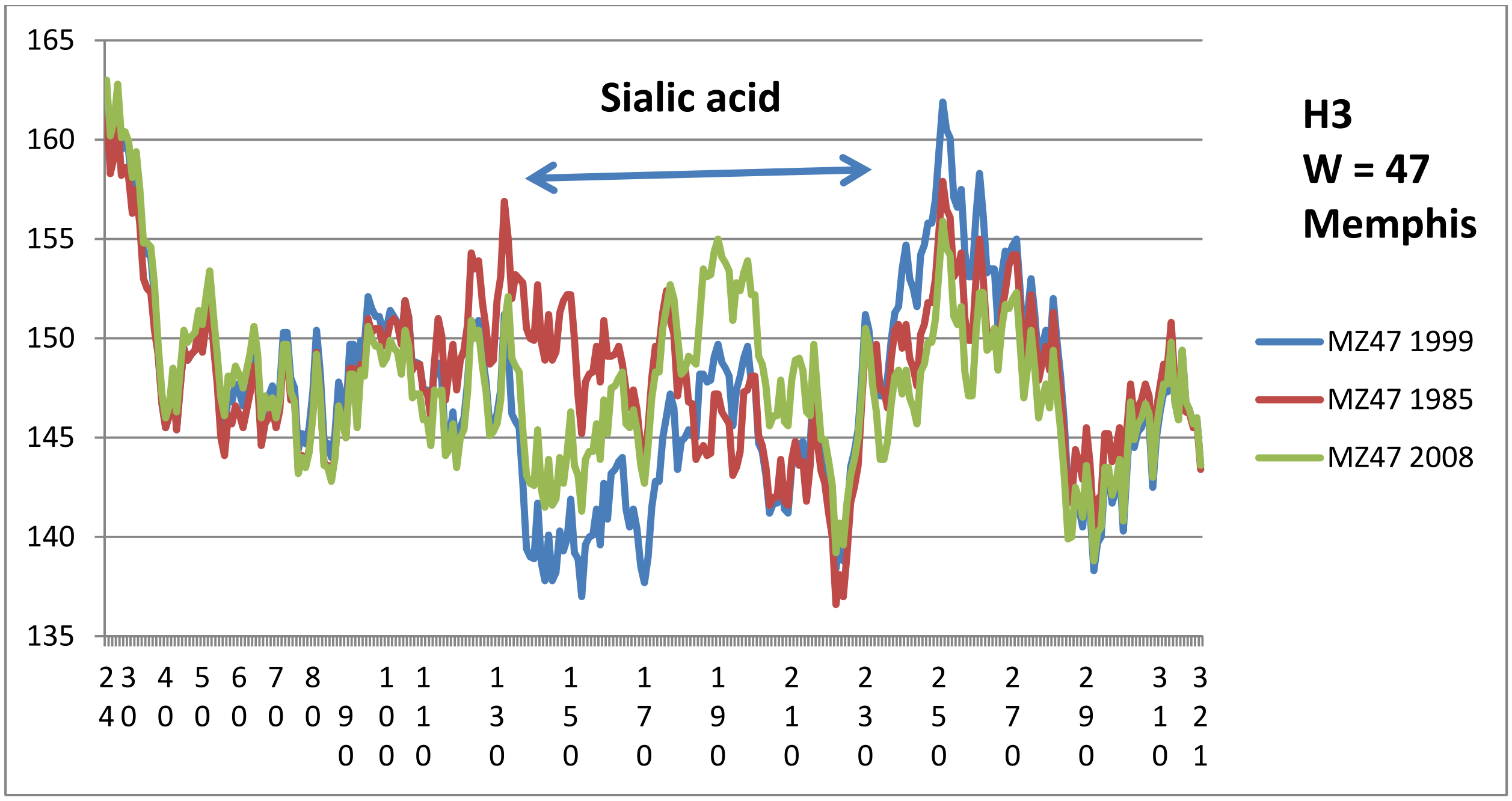

Fig. 8. There is an overall hydrophobic stiffening of H31999 into 2008 (ACD85616) in the 130-230 region, here attributed to interactions with sialic acid similar to those seen in H1 [2], but with larger effects .

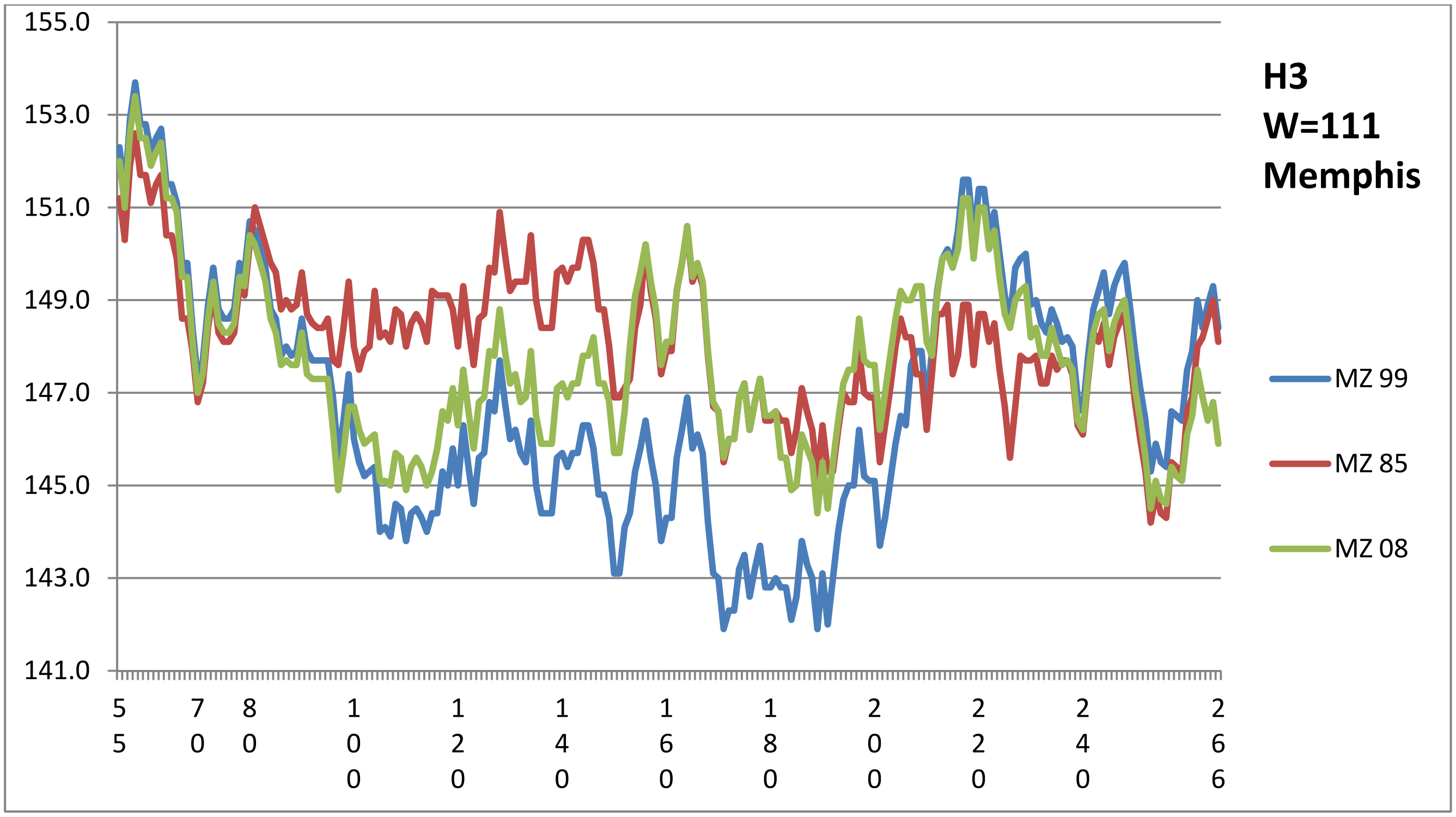

Fig. 9. Compared to R(47) in Figs. 7 and 8, here for R(111) the deep valley has narrowed and shifted to 170-190.

| | KD/<> | MZ/<> |
|---|---|---|
| A | 1.29 | 1.01 |
| C | 1.38 | 1.58 |
| D | 0.62 | 0.56 |
| E | 0.62 | 0.61 |
| F | 1.42 | 1.40 |
| G | 1.01 | 1.00 |
| H | 0.66 | 0.98 |
| I | 1.63 | 1.43 |
| K | 0.57 | 0.44 |
| L | 1.54 | 1.27 |
| M | 1.30 | 1.42 |
| N | 0.62 | 0.73 |
| P | 0.86 | 0.78 |
| Q | 0.62 | 0.68 |
| R | 0.49 | 0.50 |
| S | 0.96 | 0.64 |
| T | 0.97 | 0.87 |
| V | 1.59 | 1.53 |
| W | 0.95 | 1.12 |
| Y | 0.90 | 1.43 |

Table I. Hydropathicities according to the KD and MZ scales. To facilitate comparisons, the values have been normalized to their respective averages (<>). Note that Glycine is close to average on both scales, while Thyrosine (Y) is treated quite differently.